\documentclass{elsart}

\usepackage{natbib}

\usepackage{epsfig}

\usepackage{amssymb}

\begin{document}

\begin{frontmatter}



\title{The Phase-resolved High Energy Spectrum of the Crab Pulsar}


\author[1]{J.J. Jia},
\author[1]{Anisia P.S. Tang},
\author[2]{J. Takata},
\author[3]{H.K. Chang},
\author[1]{K.S.Cheng}

\address[1]{Department of Physics, The University of Hong Kong, Hong Kong, China, 
hrspksc@hkucc.hku.hk}
\address[2]{ASIAA/National Tsing Hua University - TIARA, PO Box 23-141, Taipei, Taiwan}
\address[3]{Department of Physics and Institute of Astronomy, National Tsing Hua 
University,
Hsinchu 30013, Taiwan}

\begin{abstract}
We present a modified outer gap model to study the phase-resolved spectra of the
Crab pulsar. A theoretical double peak profile of the light curve
containing the whole phase is shown to be consistent with the
observed light curve of the Crab pulsar by shifting the inner
boundary of the outer gap inwardly to $\sim 10$ stellar radii above
the neutron star surface. In this model, the radial distances of
the photons corresponding to different phases can be determined in
the numerical calculation. Also the local electrodynamics, such
as the accelerating electric field, the curvature radius of the
magnetic field line and the soft photon energy, are sensitive to
the radial distances to the neutron star. Using a synchrotron
self-Compton mechanism, the phase-resolved spectra with the energy
range from 100 eV to 3 GeV of the Crab pulsar can also be
explained.

\end{abstract}


\end{frontmatter}

\section{Introduction}

There are eight pulsars have been detected in gamma-ray energy range 
(cf. Thompson 2006 for a recent review) with period ranging from 0.033s to 0.237s 
and age younger than million years old. 
Theoretically, it is suggested that high-energy photons are produced by the radiation 
of charged particles that are accelerated in the pulsar magnetosphere. There are 
two kinds of theoretical models: one is the polar gap model (e.g., Harding 1981, 
Daugherty \& Harding 1996, for more detail review of polar cap model cf. Harding 2006), 
and another is the outer gap model (e.g. Cheng, Ho, \& Ruderman 1986a, 1986b; 
Chiang \& Romani 1994). Both models predict that electrons and positrons are 
accelerated in a charge depletion region called a gap by the electric field 
along the magnetic field lines and assume that
charged particles lose their energies via curvature radiation in both polar and 
outer gaps. The key differences are polar gap are located near stellar surface 
and the outer gap are located near the null charge surface, where are at least 
several tens stellar radii away the star. 

The continuous observations of powerful young pulsars, including the Crab, 
the Vela and the Geminga, have collected large number of high energy photons, which 
allow us
to carry out much more detailed analysis. Fierro et al. (1998) divided the whole 
phase into eight phase intervals, i.e. leading wing 1, peak 1, trailing 1, bridge, 
leading wing 2, peak 2, trailing 2 and off-pulse. They showed that the data in
each of these phases can be roughly fitted with a simple power law. However, the 
photon indices of these phases are very different, they range from 1.6 to 2.6. 
Massaro et al. (2000) have shown that X-ray pulse profile is energy dependent and the 
X-ray spectral index also depends on the phase of the rotation.

The Crab pulsar has been extensively studied from the radio to the extremely
high energy ranges, and the phase of the double-pulse with
separation of $144^{\circ}$ is found to be consistent over all
wavelengths. Recently Kuiper et al. (2001) have combined the X-ray and gamma-ray 
data of the Crab pulsar, they showed 
that the phase-dependent spectra exhibit a double-peak structure, i.e. 
one very broad peak in soft gamma-rays and another broad peak 
in higher energy gamma-rays. The position
of these peaks depend on the phase. Although the double-peak structure is a signature 
of synchrotron self-Compton mechanism, it is impossible to fit the phase dependent 
spectrum by a simple particle energy spectrum. 
Actually it is not surprised that the spectrum is phase dependent because photons are
emitted from different regions of the magnetosphere. The local properties, e.g. electric 
field $E(r)$, magnetic field $B(r)$, particles density and energy distribution are very 
much different for different regions. Therefore these phase dependent data 
provide very important information for emission region. Consequently, 
the phase-resolved properties provide very important clues and constraints
for the theoretical models. So far, the three-dimensional outer
gap model seems to be the most successful model in explaining both the
double-peak pulse profile and the phase-resolved spectra of the
Crab pulsar (e.g. Chiang \& Romani 1992, 1994; Dyks \& Rudak 2003;
Cheng, Ruderman \& Zhang, 2000, hereafter CRZ).
However, the leading-edge and trailing-edge of the light curve
cannot be given out, since the inner boundary of the outer gap is
located at the null charge surface in this model. Recently, the
electro-dynamics of the pulsar magnetosphere has been studied
carefully by solving the Poisson equation for electrostatic
potential and the Boltzmann equations for electrons/positrons
(Hirotani \& Shibata, 1999a,b,c; Takata et al. 2004, 2006; 
Hirotani 2005), and the inner
boundary of the gap is shown to be located near the stellar
surface.

We will organize the paper as follows. We describe the modified outer gap
model in \S 2.
In \S 3,
we calculate the phase-resolved spectra and
present the fitting result of the spectra in different phase
intervals. Finally, we discuss our results and draw conclusions in
\S 4.

\section{A modified outer gap model}

Originally proposed by Holloway (1973) that vacuum gaps may form
in the outer regions of pulsar's magnetosphere, Cheng, Ho and
Ruderman (1986a, 1986b; hereafter CHR) developed the idea of outer
magnetosphere gaps and explained the radiation mechanisms of the
$\gamma$-rays from the Crab and Vela pulsars. CHR argued that a
global current flowing through the null surface of a rapidly
spinning neutron star would result in large regions of charge
depletion, which form the gaps in the magnetosphere. They assume
the outer gap should begin at the null charge surface and extend to the light
cylinder.
In the gaps, a
large electric field parallel to the magnetic field lines is
induced ($\vec{E} \cdot \vec{B}\neq 0$), and it can accelerate the
electrons or positrons to extremely relativistic speed. Thus,
those charges can emit high energy photons through various
mechanisms, and further produce copious $e^+ e^-$ pairs to sustain
the gaps and the currents.

Based on the CHR model, Chiang and Romani (1992, 1994) generated
gamma ray light curves for various magnetosphere geometries by
assuming that gap-type regions could be supported along all field
lines which define the boundary between the closed region and open
field line region rather than just on the bundle of field lines
lying in the plane containing the rotation and magnetic dipole
axes. In their model, photons are generated and travel tangential
to the local magnetic field lines and there are beams in both the
outward and inward directions. They suggested that a single pole
would produce a double-peak emission profile when the line of
sight crosses the enhanced regions of the $\gamma$-ray beam, while
the inner region of the beam results in the bridge emission
between these two pulses. The peak phase separation can be
accommodated by choosing a proper observer viewing angle. Because
the location of emission of each point in phase along a given line
of sight can be mapped approximately in this model, the outer gap
is thus divided into small subzones. As the curvature radius,
photon densities and the local electrodynamics in different
subzones are not the same, the spectral variation of the high
energy radiation in different phase intervals varies. Later,
Romani and Yadigaroglu (1995) developed the single gap model by
involving the effects of aberration, retarded potential and time
of flight across the magnetosphere. The light curve profiles in
this modified model is simply determined by only two parameters,
which are magnetic inclination angle $\alpha$ and the viewing
angle $\zeta$. They argued that the $\gamma$-ray emission can only
be observed from pulsars with large viewing angle ($\zeta \geq
45^{\circ}$), and we cannot receive the $\gamma$ photons but radio
emissions from the aligned pulsars ($\alpha \leq 35^{\circ}$).
Furthermore, they showed the gap would grow larger as the pulsar
slows down, and more open field lines can occupy the outer gap,
which means the older pulsar are more efficient for producing GeV
$\gamma$-ray photons (Yadigaroglu \& Romani, 1995).

However, the assumptions of the model proposed by Romani's group
are not self-consistent. Why is there only a single pole and only
outgoing current in the magnetosphere? Cheng, Ruderman and Zhang
(2000) proposed another version of three dimensional outer gap
model for high energy pulsars based on the pioneering work of
Romani, and made it more natural in physics. In the CRZ model, two
outer gaps and both outward and inward currents are allowed
(though it turns out that outgoing currents dominate the emitted
radiation intensities), and the azimuthal extension of the outer
gap is restricted on a bundle of fields instead of the whole
lines. Like the previous work by Yadigaroglu and Romani (1995),
the CRZ model also contains the same two parameters, but more
self-consistent in gap geometry and radiation morphology by using
the pair production conditions. The electric field parallel to the
magnetic field lines is
\begin{eqnarray}
E_{\parallel}=\frac{\Omega B(r)f^2(r)R^2_L}{cs(r)},
\end{eqnarray}
where $f(r) \propto r^{3/2}$ and $s(r) \propto r^{1/2}$ are the fractional size 
of the outer gap and the curvature radius at the distance $r$. The characteristic 
fractional size of the outer gap evaluated at $r\sim r_L$, where
$r_L$ is the light cylinder radius, can be estimated by the condition of pair 
creation (Zhang \& Cheng 1997; CRZ) and is given by 
\begin{equation}
f_0=5.5P^{26/21}B_{12}^{-4/7}\Delta \Phi^{1/7}\;\;,
\label{size}
\end{equation}
where $\Delta \Phi$ is the
azimuthal extension  of the outer gap. CRZ
estimates its value by considering the local pair production
condition and give $\Delta \Phi \sim 160^{\circ}$ for the Crab pular. 
It has been pointed out that if the inclination angle is small, $f_0$
can be changed by a factor of several (Zhang et al. 2004). We want to 
remark that equation (1) is the solution of vacuum solution, for regions
near null surface and the inward extension of the gap the electric field
is shown to be deviated from the vacuum solution (e.g Muslimov \& 
Harding 2004; Hirotani 2006). Nevertheless for simplicity we shall assume
the vacuum solution for the entire gap.  

In the numerical calculation, the outer gap should be divided into
several layers in space. The shape of each layer at the stellar
surface is similar to that of the polar cap, but smaller in size.
Thus, for a thin gap, the calculation of only one representative
layer is enough; while for a thick one (e.g. Geminga), several
different layers should be added in the calculation (Zhang \&
Cheng, 2001). The coordinate of the footprint of the last closed
field lines on the stellar surface is determined as
$(x_0,y_0,z_0)$, then the coordinates values $(x'_0,y'_0,z'_0)$ of
the inner layers can be defined by $x'_0=a_1x_0,~y'_0=a_1y_0,$ and
$z'_0=\sqrt{1-{x'_0}^2-{y'_0}^2}$, where $a_1$ corresponds to the
various layers in the open volume.

Inside the light cylinder, high energy photons will be emitted
nearly tangent to the magnetic field lines in the corotating frame
because of the relativistic $1/\gamma$ beaming inherent in high
energy processes unless $|\bf{E}\times\bf{B}|\sim B^2$. Then the
propagation direction of each emitted photons by relativistic
charged particles can be expressed as ($\zeta$,$\Phi$), where
$\zeta$ is the polar angle from the rotation axis and $\Phi$ is
the phase of rotation of the star. Effects of the time of flight
and aberration are taken into account. A photon with velocity
${\bf{u}}=(u_x,u_y,u_z)$ along a magnetic field line with a
relativistic addition of velocity along the azimuthal angle gives
an aberrated emission direction ${\bf{u'}}=(u'_x,u'_y,u'_z)$. The
time of flight gives a change of the phase of the rotation of the
star. Combining these two effects, and choosing $\Phi=0$ for
radiation in the (x,z) plane from the center of the star, $\zeta$
and $\Phi$ are given by $\cos\zeta = u'_z$ and $\Phi
=-\phi_{u'}-{\vec{r}}\cdot{\hat{u'}}$, where $\phi_{u'}$ is the
azimuthal angle of $\hat{u'}$ and $\vec{r}$ is the emitting
location in units of $R_L$. In panel A of Fig.~1, the emission
morphology in the ($\zeta$, $\Phi$) plane is shown. For a given
observer with a fixed viewing angle $\zeta$, a double-pulsed
structure is observed because photons are clustered near two edges
of the emission pattern due to the relativistic effects (cf. panel
B of Fig.~1).

\begin{figure}[h]
\begin{center}
\includegraphics*[width=10cm]{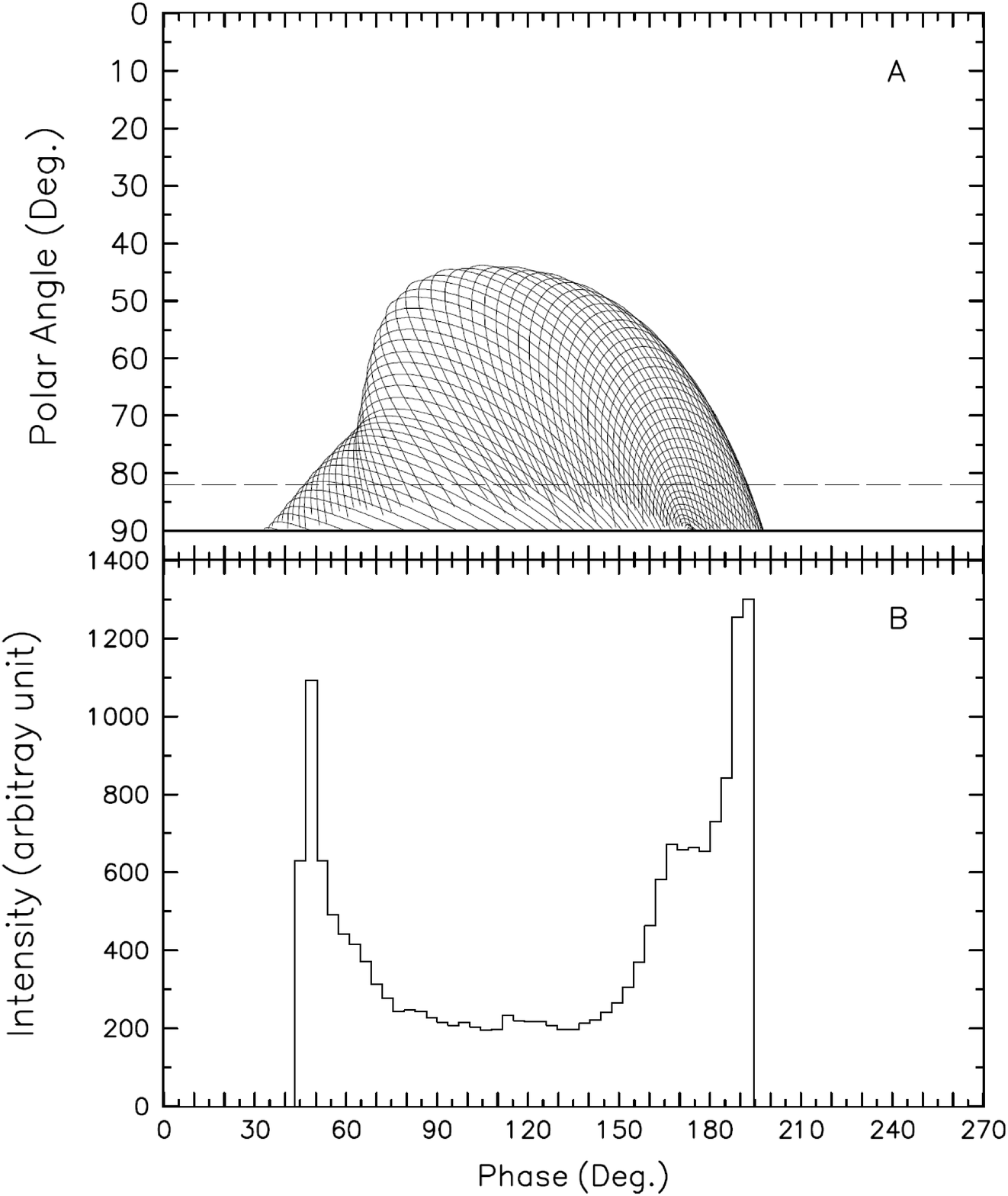}
\end{center}
\caption{Emission projection onto the ($\zeta$,$\Phi$) plane and pulse 
profile for the single pole outer gap. The photons are emitted outwards
from the outer gap. (a) The emission projection
($a1=0.9$) and (b) the corresponding pulse profile ($\Delta a1=0.03$), for
Crab parameters $\alpha=65^{\circ}$ and $\zeta=82^{\circ}$. }
\label{fig:1}
\end{figure}

In Fig.~1, we can see that this model can only produce radiation between
two peaks. However, the observed data of the Crab, Vela and Geminga 
indicate that the leading wing 1 and the trailing wing 2 are quite strong, and even 
the intensity in off-pulse cannot be ignored. Hirotani and his co-workers 
(Hirotani \& Shibata 2001; Hirotani, Harding \& Shibata 2003) have pointed out that the 
large current in the outer gap can change the boundary of the outer gap. They solve the 
set of Maxwell and Boltzmann equations in pulsar magnetospheres and demonstrate the 
existence of outer-gap accelerators, whose inner boundary position depends the
detail of the current flow and it is not necessarily located at the null 
charge surface. For the gap current lower than 25\% of the Goldreich-Julian current,
the inner boundary of the outer gap is very close to the null surface (Hirotani 2005). 
On the other hand if the current is close to the Goldreich-Julian current, the inner 
boundary can be as close as 10 stellar radii. 
In Fig.~2, we show the light curve by assuming the inner boundary is extended inward 
from the null charge surface to 10 stellar radii (cf. panel A of Fig.~2). 
In panel B of Fig.~2, the solid line represents emission trajectory of 
outgoing radiation of one gap from the null surface to the light cylinder  
with $\alpha$ = 50$^\circ$ and 
$\zeta =75^\circ$ and the dashed line represents the outgoing radiation from 
another gap from the inner boundary to the null surface. In the presence of the 
extended emission region from the near the stellar surface to the null charge 
surface, leading wing 1, trailing wing 2 and the off-pulse components can 
also be produced.

\begin{figure}[h]
\begin{center}
\includegraphics*[width=10cm]{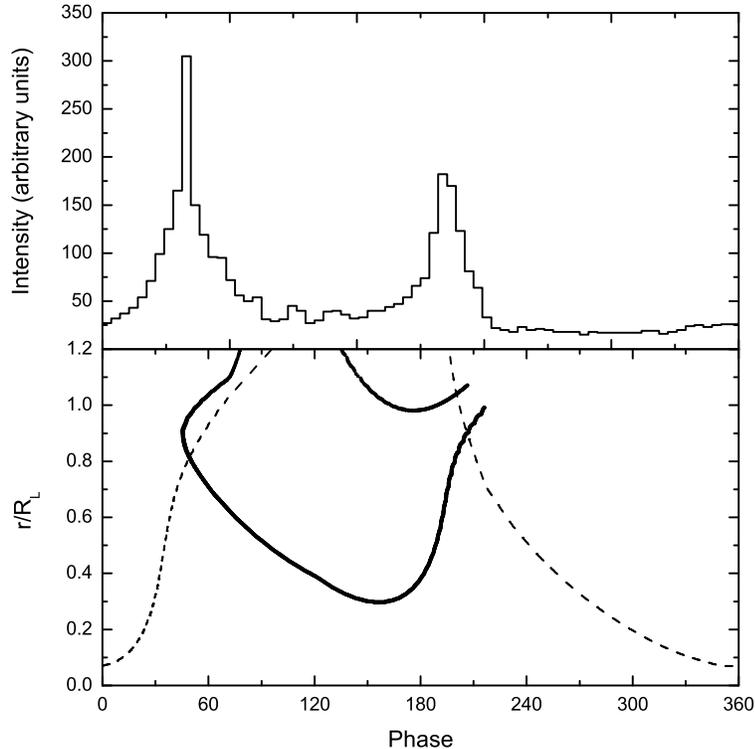}
\end{center}
\caption{Upper panel: the simulated pulse profile of the Crab pulsar;
 lower panel: variation of radial distance with pulse phase for the Crab pulsar in units
 of $R_L$, where the bold line represents those in the outer
 magnetosphere, and the dashed line represents those in the inner
 magnetosphere. The inclination angle is $50^{\circ}$ and the viewing angle is 
$75^{\circ}$.}
\label{fig:2}
\end{figure}

\section{The phase-resolved spectra}

\subsection{radiation spectrum}
The Crab pulsar has enough photons for its spectra to be analyzed,
and the phase-resolved spectra are useful for study of the local
properties of the magnetosphere. Here, we summarize the calculation 
procedure of the radiation spectrum given in CRZ, which is used to calculate 
the spectrum in different phases.

The electric field of a thin outer gap (CHR) is given by 
$E_{\parallel}(r)=\frac{\Omega B(r)a^2(r)}{cs(r)}=\frac{\Omega
B(r)f^2(r)R^2_L}{cs(r)}$,
where $a(r)$ is the thickness of the outer gap at position $r$,
and $f(r)=a(r)/R_L$ is the local fractional size of the outer gap.
Assuming that the magnetic flux subtended in the outer gap is
constant in the steady state, we get the local size factor
$f(r)\sim f(R_L)(\frac{r}{R_L})^{3/2}$,
where $f(R_L)$ is estimated by using the pair creation condition (cf. 
Zhang \& Cheng 1997, CRZ). 
As the equilibrium between the
energy loss in radiation and gain from accelerating electric
field, the local Lorentz factor of the electrons/positrons in the
outer gap is $\gamma_e(r)=(\frac{3}{2}\frac{s^2}{e^2c}eE_{\parallel}(r)c)^{1/4}$.

For a volume element $\Delta V$ in the outer gap, the number of
primary charged particles can be roughly written as
$dN=n_{\rm{GJ}}\Delta A\Delta l$,
where $n_{\rm{GJ}}=\frac{\bf{\Omega \cdot B}}{2\pi ec}$ is the
local Goldreich-Julian number density, $B\Delta A$ is the magnetic
flux through the accelerator and $\Delta l$ is the path length
along its magnetic field lines. (Here, We would like to remark that 
this could overestimate the primary charge number density near the null
surface, where the positronic charge density dominates the Goldreich-Julian 
charge density. However, the observed radiation comes from the wide range of 
magnetospheric region, an slight overestimation of a small region should 
not cause a qualitative difference.)
Thus, the total number of the
charged particles in the outer gap is
$N\sim \frac{\Omega \Phi}{4\pi ce}R_L$,
where $\Phi \sim f(R_L)B(R_L)R^2_L\Delta \phi$ is the typical angular
width of the magnetic flux tube subtend in the outer gap.
The primary $e^{\pm}$ pairs radiate curvature photons with a
characteristic energy $E_{cur}(r)=\frac{3}{2}\hbar \gamma^3_e(r)\frac{c}{s(r)}$,
and the power into curvature radiation for $dN$ $e^{\pm}$ pairs in
a unit volume is $\frac{dL_{cur}}{dV}\approx l_{cur}n_{\rm{GJ}}(r)$,
where $l_{cur}=eE_{\parallel}c$ is the local power into the
curvature radiation from a single electron/positron. The spectrum
of primary photons from a unit volume is
\begin{eqnarray}
\frac{d^2\dot{N}}{dVdE_{\gamma}}\approx
\frac{l_{cur}n_{\rm{GJ}}}{E_{cur}}\frac{1}{E_{\gamma}},~~E_{\gamma}\leq
E_{cur}.
\end{eqnarray}
These primary curvature photons collide with the soft photons
produced by synchrotron radiation of the secondary $e^{\pm}$
pairs, and produce the secondary $e^{\pm}$ pairs by photon-photon
production. In a steady state, the distribution of secondary
electrons/positrons in a unit volume is
\begin{eqnarray}
\frac{d^2N}{dVdE_e}\approx \frac{1}{\dot{E}_e}\int
\frac{d^2\dot{N}(E'_{\gamma}=2E'_e)}{dVdE_{\gamma}}dE'_e\approx
\frac{1}{\dot{E}_e}\frac{l_{cur}n_{\rm{GJ}}}{E_{cur}}\ln(\frac{E_{cur}}{E_e}),
\end{eqnarray}
with $\dot{E}_e$ the electron energy loss into synchrotron
radiation, which is
$\dot{E}_e=-\frac{2}{3}\frac{e^4B^2(r)\sin^2\beta(r)}{m^2_ec^3}(\frac{E_e}{m_ec^2})^2$,
where $B(r)$ is the local magnetic field and $\beta(r)$ the local
pitch angle,
$\sin \beta(r)\sim \sin \beta(R_L)(\frac{r}{R_L})^{1/2}$,
$\sin \beta(R_L)$ is the pitch angle at the light cylinder.
Therefore, the energy distribution of the secondary
electrons/positrons in volume $\Delta V(r)$ can be written as
\begin{eqnarray}
\frac{dN(r)}{dE_e}\approx \frac{d^2N}{dVdE_e}\Delta V(r)\sim
\frac{1}{\dot{E}_e}\frac{l_{cur}n_{\rm{GJ}}\Delta
V(r)}{E_{cur}}\ln (\frac{E_{cur}}{E_e}).
\end{eqnarray}
The corresponding photon spectrum of the synchrotron radiation is
\begin{eqnarray}
F_{syn}(E_{\gamma},r)=\frac{\sqrt{3}e^3B(r)\sin
\beta}{m_ec^2h}\frac{1}{E_{\gamma}}\int^{E_{max}}_{E_{min}}\frac{dN(r)}{dE_e}F(x)dE_e,
\end{eqnarray}
where $x=E_{\gamma}/E_{syn}$, and
$E_{syn}(r)=\frac{3}{2}(\frac{E_e}{m_ec^2})^2\frac{h e B(r)\sin
\beta(r)}{m_ec}$
is the typical photon energy, and $F(x)=x\int^{\infty}_x
K_{5/3}(y)dy$, where $K_{5/3}(y)$ is the modified Bessel function
of order 5/3. Also, the spectrum of the inverse Compton scattered
photons in the volume $\Delta V(r)$ is
\begin{eqnarray}
F_{ICS}(E_{\gamma},r)=\int^{E_{max}}_{E_{min}}\frac{dN(r)}{dE_e}\frac{d^2N_{ICS}(r)}{dE_{
\gamma}dt}dE_e,
\end{eqnarray}
where
$\frac{d^2N_{ICS}(r)}{dE_{\gamma}dt}=\int^{\epsilon_2}_{\epsilon_1}n_{syn}(\epsilon,r)F(\%
epsilon,E_{\gamma},E_e)d\epsilon$,
and
$F(\epsilon,E_{\gamma},E_e)=\frac{3\sigma_Tc}{4(E_e/mc^2)^2}\frac{1}{\epsilon}[2q\ln
q+(1+2q)(1-q)+\frac{(\Gamma q)^2(1-q)}{2(1+\Gamma q)}]$,
where $\Gamma=4\epsilon (E_e/m_ec^2)/m_ec^2$, $q=E_1/\Gamma
(1-E_1)$ with $E_1=E_{\gamma}/E_e$ and $1/4(E_e/m_ec^2)<q<1$. The
number density of the synchrotron photons with energy $\epsilon$
is $n_{syn}(\epsilon,r)=\frac{F_{syn}(\epsilon)}{cr^2\Delta \Omega}$,
where $F_{syn}$ is the calculated synchrotron radiation flux, and
$\Delta \Omega$ is the usual beam solid angle.

\begin{figure}[h]
\begin{center}
\includegraphics*[width=17cm,angle=90]{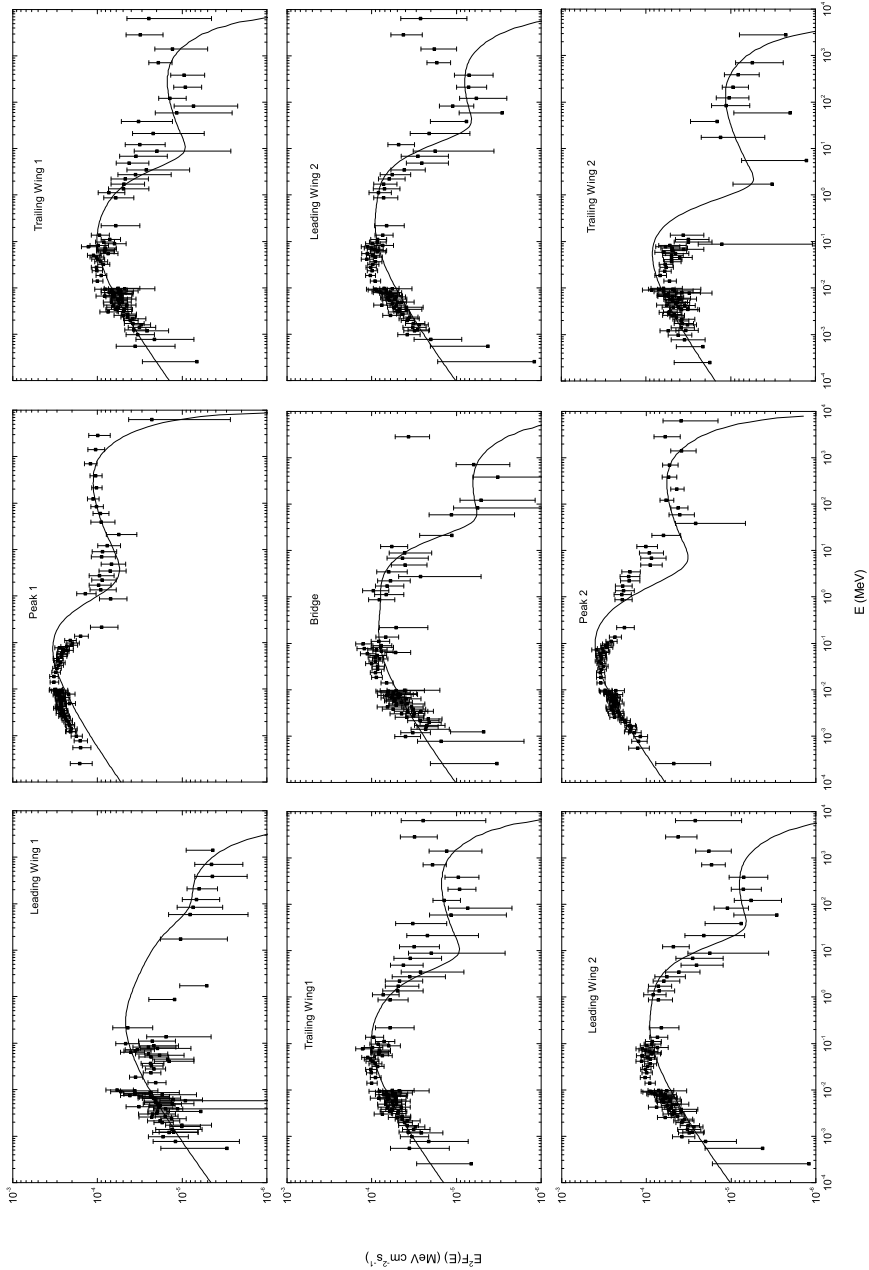}
\end{center}
\caption{Phase resolved spectra of the Crab pulsar from 100 eV to 3 GeV in the 7 narrow
 pulse-phase intervals. Two spectra (for the TW1 and LW2) are displayed twice. The curved 
line is
 calculated by the theoretical model, and the observed data are taken from Kuiper et al. 
(2001).}
\label{fig:3}
\end{figure}

Fig.~3 shows the observed data of the phase-resolved spectra from
100 eV to 3 GeV of the Crab pulsar, and the theoretical fitting
results calculated by using the synchrotron self-Compton
mechanism. The phase intervals are defined by division given by
Fierro (1998), and the amplitude of the spectrum in each phase
interval is proportional to the number of photons counted in it.
In this fitting, $f(R_L)=0.21$, and $B=3.0\times
10^{12}\rm{Gauss}$ are used, which give a consistent fitting of
the phase-resolved spectra of the seven phase intervals. In order to 
obtain a better fit, we treat the pitch angle ($\beta$) and the beam solid
angle ($\Delta \Omega$) near the light cylinder as free parameters
and vary from phase to phase in the calculation. $\sin
\beta(R_L)=0.06$ and $\Delta \Omega=5.0$ are chosen for trailing
wing 1, bridge and leading wing 2; $\sin \beta(R_L)=0.02, \Delta
\Omega=1.0$ for leading wing 1; $\sin \beta(R_L)=0.04, \Delta
\Omega=3.5$ for peak 1, $\sin \beta(R_L)=0.07, \Delta \Omega=3.0$
for peak 2, and $\sin \beta(R_L)=0.03, \Delta \Omega=6.0$ for
trailing wing 2.
Additionally, the phase-averaged spectrum of the total pulse of
the Crab pulsar is shown in Fig.~4, where the parameters are
chosen as $\sin \beta(R_L)=0.05$ and $\Delta \Omega=5.0$.

\begin{figure}[h]
\begin{center}
\includegraphics*[width=10cm]{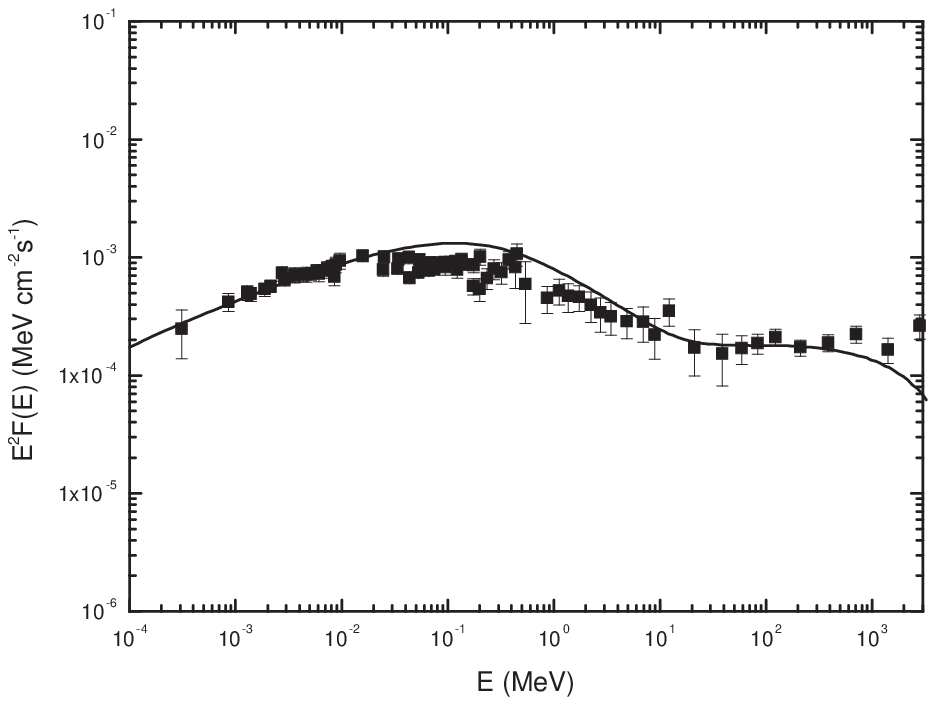}
\end{center}
\caption{Phase-averaged spectrum of the Crab pulsar. The observed data are taken from 
Kuiper et al. (2001).}
\label{fig:4}
\end{figure}

\subsection{Analysis of the Phase-Resolved Spectra}

The high energy spectra of the Crab pulsar is explained by using
the synchrotron self-Compton mechanism, which involves both the
synchrotron radiation and the Inverse Compton-Scattering (ICS)
caused by the ultra-relativistic electron/positron pairs created
by the extremely high-energy curvature photons. The secondary
$e^{\pm}$ pairs gyrate in the strong magnetic field and radiate
synchrotron photons. While in the far regions of the magnetosphere
where the magnetic field decays rapidly, the relativistic pairs
collide with the soft synchrotron photons through the ICS process.
Thus, the spectra of the radiation contain two main components:
one is the synchrotron radiation from the soft X-ray to $\sim$10
MeV, and the other is the ICS component in the even higher energy
range. Usually, the synchrotron spectrum has stronger amplitude
than that of ICS, and there is obvious turning frequency between
these two components, e.g. about 3MeV for peak 1. As we know, the
power of synchrotron radiation and ICS can be compared by the
ratio of the local magnetic energy density and the photon energy
density, i.e.
\begin{eqnarray}
\frac{P_{syn}}{P_{ICS}}\approx \frac{U_{B}}{U_{ph}}\propto
\frac{B(r)^2/8\pi}{\epsilon_{syn}(r)n_{syn}(\epsilon_{syn},r)},
\end{eqnarray}
where $\epsilon_{syn}(r)$ is the synchrotron photon energy in
location $r$. In Fig.~3, the spectra in trailing wing 1, bridge
and leading wing 2 have broad synchrotron spectra, which cover
from 100 eV to $\sim$30 MeV. In Fig.~2, it is demonstrated that
the radiation of these three phase intervals are dominated by the
photons generated in the near surface region, where the magnetic
field is so strong that synchrotron radiation takes up the most
emission. However, the radiation of peak 1 and 2 are from the far
regions near the light cylinder, where the magnetic field decays
rapidly $(B\propto r^{-3})$, thus, the ICS radiation becomes more
important above 3 MeV.

The peak of the synchrotron spectrum is determined by the
characteristic synchrotron photon energy. Since
$E_{syn}\propto \gamma^2_{e}B\sin \beta$, where $\gamma_e$ is the
Lorentz factor of the secondary pairs and $\beta$ is the pitch
angle of the electron/positron to the magnetic field, the peak of
the synchrotron spectrum can shift if the $\beta$ varies. Since
the outward radiation direction covers a wider range than that of the inward radiation,
so the solid angle $(\Delta \Omega)$ is no longer the unity as
assumed in CRZ model. The solid angle
can effect the amplitude of the ICS spectrum because the number
density of the synchrotron photons is proportional to
$\frac{1}{\Delta \Omega}$. Therefore it is
reasonable for us to choose $\beta (R_L)$ and $\Delta \Omega$ as a
set of parameters in fitting the phase-resolved spectra of the
Crab pulsar.

\section{Conclusion and Discussion}

We have tried to explain the high energy light curve and the
phase-resolved spectra in the energy range from 100 eV to 3 GeV of
the Crab pulsar by modifying the three dimensional outer
magnetosphere gap model. Compared to the classical outer gap with
the inner boundary at the null charge surface, the modified model
allows the outer gap to start at the region about several stellar
radii above the neutron star surface, and the "inwardly-extended"
part of the outer gap contributes to the outer wings and off-pulse
of the light curve. Such modified outer gap
geometry also plays a vital role in explaining the optical 
polarization properties of the Crab pulsar (Takata et al. 2006).
Two adjustable parameters are used to simulate
the light curve: one is the inclination angle of the magnetic axis
to the rotational axis $\alpha$, and the other is the viewing
angle also to the rotational axis $\zeta$. As constrained by the
phase separation of the double peaks, we choose the values for
these two parameters that $\alpha=50^{\circ}$ and
$\zeta=75^{\circ}$. So far, these two parameters have not been
determined from the observations. From radio observations, Rankin
(1993) estimated that $\alpha \approx 84^{\circ}$ and $\zeta$ is
not known. Moffett and Hankins (1999) gave that $\alpha \approx
56^{\circ}$ and $\zeta=117^{\circ}$ by using the polarimetric
observations at frequencies between 1.4 and 8.4 GHz. Of course,
our values cannot be the true ones, and require further
observations to give strong restrictions of them.

In fitting the phase-resolved spectra of the Crab pulsar, our
model performs well from 100 eV to 1 GeV, but fails beyond 1 GeV.
The inverse Compton scattering spectrum of our results falls down
quickly when the energy is over 1 GeV, but the observation data
indicates that the spectrum still increases, especially in the
first trailing wing, the bridge and the second leading wing phase
intervals. We have assumed that the curvature photons are all absorbed by the
magnetic field lines, however, some of these multi-GeV
photons produced near the light cylinder should be easily escaped
from the photon-photon pair creation process. In the spectrum fitting of peak 1, our
result has a frequency shift below 1 MeV, and we found that in
order to well fit the spectrum we should reduce the curvature
photon energy by a quarter. The energy of the curvature photon
$E_{cur}\propto s^{-1}(r)$, where $s(r)$ is the local curvature
radius. As the high energy photons are produced in the far regions
of the magnetosphere, where $s(r)$ maybe not follow the dipole
form, we can change the photon energy slightly.

Moreover, the stellar radius of a neutron star is usually treated
as $10^6$ cm when calculating the strength of the surface magnetic
field. However, the equation of state inside the neutron star of
the current theoretical models cannot give a convincing value of
the neutron star size. Thus, we can only determine the magnetic
moment, i.e. $B_pR^3_0$, of the pulsar from the energy loss rate.
Therefore, we can rewrite the magnetic field of the Crab pulsar as
$B_{12}R^3_{6}=3.8$. In fitting the phase-resolved spectra of the
Crab pulsar, we choose $B_p=3\times 10^{12}$ G, not the
traditional value of $3.8\times 10^{12}$ G, for it gives a better
fitting on the lower energy range below 10 keV.

Finally we want to emphasize that 


This research is supported by a RGC grant of Hong Kong Government under
HKU 7015/05P.




\end{document}